\documentclass{elsart}
\usepackage{epsfig,amsmath,amscd,amssymb}
\begin{document}
\begin{frontmatter}

\begin{flushright}
{INFNFE-06-01}
\end{flushright}

\title{High-energy neutrino oscillations in absorbing matter}
\author{Vadim A. Naumov\thanksref{emVN}\thanksref{Supp}}
 \address{Dipartimento di Fisica and Sezione INFN di Ferrara,
         Via del Paradiso 12, I-44100 Ferrara, Italy}
 \address{Laboratory for Theoretical Physics, Irkutsk State
         University, Gagarin boulevard 20, RU-664003 Irkutsk,
         Russia}
\thanks[emVN]{E-mail: naumov@fe.infn.it}
\thanks[Supp]{Partially supported by grant from Ministry of Education
              of Russian Federation, within the framework of the
              program ``Universities of Russia -- Basic Researches''.}
\begin{abstract}
The impact of neutrino mixing, refraction and absorption on
high-energy neutrino propagation through a thick medium is studied
using the MSW evolution equation with complex indices of refraction.
It is found that, owing to the mixing with sterile
neutrinos, the penetrability of active neutrinos may be many orders
of magnitude larger than it would be in the absence of mixing.
The effect is highly sensitive to changes in density and composition
of the matter background as well as to neutrino energy and mixing
parameters. This may lead to observational consequences in neutrino
astrophysics.
\end{abstract}
\begin{keyword}
Neutrino oscillations \sep Mixing \sep Refraction \sep Absorption
\sep Sterile neutrinos
\PACS 14.60.Lm \sep 14.60.Pq \sep 13.15.+g \sep
      14.60.St \sep 96.40.Tv \sep 05.60.+w
\end{keyword}
\end{frontmatter}

\clearpage
\section{Introduction}
\label{sec:Introduction}

In studies of high-energy neutrino propagation through matter two
completely different approaches are in use: the quantum-mechanical
MSW formalism \cite{MSW} and the classical transport theory
\cite{Berezinsky86}.
The first approach accounts for the neutrino mixing and coherent
interference of propagating neutrinos (that is refraction) neglecting
the loss of coherence due to inelastic scattering. The second one
provides the means for estimating neutrino absorption, production and
energy loss in inelastic charged and neutral current interactions,
but is inapplicable for the inclusion of quantum effects of neutrino
mixing and refraction.
In other words, the necessary conditions for the validity of
these approaches are mutually exclusive.
At the same time there is a multitude of physically feasible
and interesting contexts where the effects of neutrino mixing,
refraction and inelastic collisions may be equally important or
comparable.
Well-known examples are the propagation of high-energy neutrinos
produced in the sun from cosmic ray interactions and from
annihilations of heavy dark matter particles captured by the sun
(see, e.g., ref. \cite{Examples} and references therein).
The so-called hidden sources (the astrophysical neutrino sources
opaque for electromagnetic radiation) \cite{HiddenSources}
hold promise as the splendid ``testing laboratories'' for the
effects under discussion because some types of these objects
(e.g., a Thorne--\.Zytkow star \cite{Thorne77}) have relevant
dimensions and density profiles.

The necessity of a unified description of mixed neutrinos
in hot media (like the $CP$ symmetric plasma of the early Universe,
a supernova core or a protoneutron star) has long been realized
\cite{CosmoNu} and the rigorous quantum kinetic theory has been
developed \cite{KineticTheory}.
\footnote{For further examples of extensive study of the interplay
          between neutrino oscillations and neutrino scattering
          see ref. \cite{RelicNeutrinos}.
          Another closely related topic is a combined treatment of
          neutrino mixing and decay \cite{OscillationsPlusDecay}.}
However, the direct application of that theory to the problem of
high-energy neutrino transport in a normal cold matter is very
involved and the analytic treatment of the quantum kinetic
equations \cite{KineticTheory} is hardly possible, except for
very particular cases.

Since an analytic analysis is always useful to gain an insight
into a sophisticated problem, it seems instructive to apply the
conventional MSW formalism straightforwardly generalized by
incorporating the imaginary parts into the neutrino indices of
refraction.
This will allow an understanding of the impact of the neutrino
absorption in conjunction with mixing and refraction, still
disregarding the neutrino energy loss from the neutral current
interactions \cite{Berezinsky86,NeutralCurrentEffect},
charged current induced chains like
$\nu_\tau N\to\tau X$, $\tau\to\nu_\tau X'$ \cite{NuTauReg}
and neutrino flavor changing reactions
$\overline{\nu}_ee^-\to\overline{\nu}_\mu\mu^-$,
$\nu_\mu e^-\to\nu_e\mu^-$, etc.
\cite{NeutrinoInteractions}.
The present paper is concerned with some surprising phenomena which
follow from this simple approach. For further simplification we
limit ourselves to studying two-flavor case. Some general results
relevant to the subject were discussed previously in
ref. \cite{Korenblit95}.

\protect\section{Master equation}
\label{sec:ME}

The generalized MSW equation for the time-evolution operator
\[
\mathbf{S}(t)=
\begin{pmatrix}
S_{\alpha\alpha}(t) & S_{\alpha\beta }(t) \\
S_{\beta \alpha}(t) & S_{\beta \beta }(t)
\end{pmatrix}
\]
of two mixed stable neutrino flavors $\nu_\alpha$ and
$\nu_\beta$ propagating through a matter with absorption can be written as
\begin{equation}\label{WE}
i\frac{\d}{\d t}\mathbf{S}(t)=\left[\mathbf{V}\mathbf{H}_0\mathbf{V}^T
+\mathbf{W}(t)\right]\mathbf{S}(t), \quad
\left(\mathbf{S}(0)=\mathbf{1}\right).
\end{equation}
Here
\[
\mathbf{V}=
\begin{pmatrix}
 \cos\theta & \sin\theta \\
-\sin\theta & \cos\theta
\end{pmatrix}
\]
is the vacuum mixing matrix that relates flavor to mass eigenstates,
$\theta$ is the mixing angle ($0\leq\theta\leq\pi/2$),
\begin{equation}\label{H0andW}
\mathbf{H}_0=
\begin{pmatrix}
E_1 &  0  \\
 0  & E_2
\end{pmatrix},
\quad
\mathbf{W}(t)=-p_\nu
\begin{pmatrix}
n_\alpha(t)-1 & 0 \\
0 & n_\beta(t)-1
\end{pmatrix},
\end{equation}
$E_i=\sqrt{p_\nu^2+m_i^2}\simeq p_\nu+m_i^2/2p_\nu$ and $m_i$ are
the energies and masses of the neutrino mass eigenstates,
respectively,%
\footnote{We assume as usual that
          $p_\nu^2 \simeq E_\nu^2\gg\max\left(m_i^2\right)$.}
$n_\alpha(t)$ is the complex index of refraction,
\[
n_\alpha(t)=1+\frac{2\pi N_0\rho(t)}{p_\nu^2}
\sum_kY_k(t)f_{\nu_{\alpha}k}(0),
\]
where $N_0=6.022\times10^{23}\,\text{cm}^{-3}$, $f_{\nu_{\alpha}k}(0)$ is
the amplitude for the $\nu_\alpha$ zero-angle scattering from particle
$k$ ($k=e,p,n,\ldots$), $\rho(t)$ is the density of the matter
(in g/cm${}^3$) and $Y_k(t)$ is the number of particles $k$
per amu in the point $t$ of the medium.
From the optical theorem (see, e.g., ref. \cite{Goldberger67})
we have $\text{Im}\left[f_{\nu_{\alpha}k}(0)\right]=(p_\nu/4\pi)
\sigma_{\nu_{\alpha}k}^{\text{tot}}\left(p_\nu\right)$,
where $\sigma_{\nu_{\alpha}k}^{\text{tot}}\left(p_\nu\right)$
is the total cross section for $\nu_{\alpha}k$ scattering.
This implies that
\begin{equation}\label{Imn}
p_\nu\text{Im}\left[n_{\alpha}(t)\right]=\frac{N_0\rho(t)}{2}
\sum_kY_k(t)\sigma_{\nu_{\alpha}k}^{\text{tot}}\left(p_\nu\right)
=\frac{1}{2\varLambda_\alpha(t)},
\end{equation}
where $\varLambda_{\alpha}(t)$ is the mean free path of neutrino
$\nu_\alpha$ in the point $t$ of the medium.

It is useful to transform eq. \eqref{WE} into the one with a
traceless Hamiltonian. For this purpose we define the matrix
\begin{equation}\label{NewS}
\widetilde{\mathbf{S}}(t)=\exp\left\{\frac{i}{2}\int_0^t\text{Tr}
\left[\mathbf{H}_0+\mathbf{W}(t')\right]\d t'\right\}\mathbf{S}(t).
\end{equation}
After substituting eq. \eqref{NewS} into eq. \eqref{WE}, we have
\begin{equation}\label{NewWE}
i\frac{\d}{\d t}\widetilde{\mathbf{S}}(t)=
\mathbf{H}(t)\widetilde{\mathbf{S}}(t), \quad
\widetilde{\mathbf{S}}(0)=\mathbf{1}.
\end{equation}
Here
\begin{equation}\label{NewH}
\mathbf{H}(t)=
\begin{pmatrix}
q(t)-\Delta_c & \Delta_s \\
\Delta_s      & -q(t)+\Delta_c
\end{pmatrix},
\end{equation}
\begin{gather*}
\Delta_c=\Delta\cos2\theta, \quad
\Delta_s=\Delta\sin2\theta, \quad
\Delta=\frac{m_2^2-m_1^2}{4p_\nu}, \\
q(t)=q_R(t)+iq_I(t)=\frac{1}{2}
p_\nu\left[n_\beta(t)-n_\alpha(t)\right].
\end{gather*}
The Hamiltonian for antineutrinos is of the same form
as eq. \eqref{NewH} but one must keep in mind that
$\text{Re}\left[f_{\overline{\nu}_{\alpha}k}(0)\right]=
-\text{Re}\left[f_{\nu_{\alpha}k}(0)\right]$, while
$\sigma_{\nu_{\alpha}k}^{\text{tot}}(p_\nu)$ and
$\sigma_{\overline{\nu}_{\alpha}k}^{\text{tot}}(p_\nu)$
are different in magnitude even at very high energies.

The neutrino oscillation probabilities are just the
squared absolute values of the elements of the evolution
operator $\mathbf{S}(t)$,
\begin{equation}\label{ProbDef}
P\left[\nu_\alpha(0)\to\nu_{\alpha'}(t)\right]\equiv
P_{\alpha\alpha'}(t)=\left|S_{\alpha'\alpha}(t)\right|^2.
\end{equation}
Taking into account eqs. \eqref{H0andW},
\eqref{Imn}, \eqref{NewS} and \eqref{ProbDef} yields
\begin{equation}\label{ProbNew}
P_{\alpha\alpha'}(t)=A(t)
\left|\widetilde{S}_{\alpha'\alpha}(t)\right|^2,
\end{equation}
where
\[
A(t)=\exp\left[-\int_0^t\frac{\d t'}{\varLambda(t')}\right], \quad
\frac{1}{\varLambda(t)}=
 \frac{1}{2}\left[\frac{1}{\varLambda_\alpha(t)}
+\frac{1}{\varLambda_\beta(t)}\right].
\]

Owing to the complex potential $q$, the Hamiltonian \eqref{NewH}
is non-Hermitian and the new evolution operator \eqref{NewS}
is nonunitary. As a result, there are no conventional relations
between the four probabilities given by eq. \eqref{ProbNew}.
Since
\[
q_I(t)=\frac{1}{4}\left[\frac{1}{\varLambda_\beta(t)}
                       -\frac{1}{\varLambda_\alpha(t)}\right],
\]
the matrix $\mathbf{H}(t)$ becomes Hermitian when
$\varLambda_\alpha=\varLambda_\beta$. If this is the case at
any $t$, eq. \eqref{NewWE} reduces to the standard MSW equation
and inelastic scattering in the medium results in the common
exponential attenuation of the survival and transition
probabilities.
From here, we shall consider the more general and more
interesting case, when $\varLambda_\alpha\neq\varLambda_\beta$.

The extreme example of such a case is provided by mixing
between the ordinary ($\alpha=e,\mu$ or $\tau$) and sterile
($\beta=s$) neutrino flavors. Since $\varLambda_s=\infty$,
we have $\varLambda=2\varLambda_\alpha$ and
$q_I=-1/4\varLambda_\alpha$. So $q_I$ is nonzero at any energy.
Even without solving the evolution equation, one can expect the
penetrability of active neutrinos to be essentially modified in
this case because, roughly speaking, they spend a certain part
of life in the sterile state.

Other important examples are the $\nu_e-\nu_\tau$ and
$\nu_\mu-\nu_\tau$ mixings. The total cross sections
for $e$ or $\mu$ production are well above the one for
$\tau$ production within a wide energy range \cite{TauAppearance}.
This is because of large value of $\tau$ lepton mass $m_\tau$
which leads to high thresholds and sharp shrinkage of the
phase spaces for the charged current $\nu_\tau N$ reactions
as well as to the kinematic correction factors ($\propto m_\tau^2$)
to the nucleon structure functions; the corresponding structures
are negligible in the case of electron production and small for
muon production.
The neutral current contributions are, of course, cancelled out
from $q_I$. Thus, in the context of the master equation
\eqref{NewWE}, $\nu_\tau$ can be treated as (almost) sterile
within the energy range for which
$\sigma_{\nu_{e,\mu}N}^{\text{CC}}\gg
\sigma_{\nu_{\tau}N}^{\text{CC}}$.

A similar situation, while in quite a different and narrow energy
range, holds in the case of mixing of $\overline{\nu}_e$ with some
other flavor. This is a particular case for a normal $C$ asymmetric
medium, because of the $W$ boson resonance formed in the
neighborhood of $E_\nu^{\text{res}}\approx6.33$ PeV through the
reactions
\[
\overline{\nu}_ee^-\to W^-\to\text{hadrons}
\quad\text{and}\quad
\overline{\nu}_ee^-\to W^-\to\overline{\nu}_\ell\ell^-
\quad(\ell=e,\mu,\tau).
\]
Just at the resonance peak,
$\sigma_{\overline{\nu}_ee}^{\text{tot}}\approx
250\,\sigma_{\overline{\nu}_eN}^{\text{tot}}$
(see ref. \cite{NeutrinoInteractions} for further details and
references).

For $E_\nu\ll\text{min}\left(m_{W,Z}^2/2m_k\right)$
(where $m_{k,W,Z}$ denote the masses of the scatterers $k$ and
gauge bosons) and for an electroneutral nonpolarized cold medium,
the real part of the potential $q$ is energy independent.
In the leading orders of the standard electroweak theory it is
\cite{RefractiveIndex}
\begin{equation}\label{qR_Examples}
q_R=\begin{cases}
\tfrac{1}{2}V_0Y_p\rho
                & \text{for $\alpha=e$ and $\beta=\mu$ or $\tau$}, \\
-\tfrac{1}{2}a_\tau V_0\left( Y_p+b_\tau Y_n\right)\rho
                & \text{for $\alpha=\mu$ and $\beta=\tau$},        \\
\tfrac{1}{2}V_0\left(Y_p-\tfrac{1}{2}Y_n\right)\rho
                & \text{for $\alpha=e$ and $\beta=s$},             \\
\tfrac{1}{4}V_0Y_n\rho
                & \text{for $\alpha=\mu$ or $\tau$ and $\beta=s$},
\end{cases}
\end{equation}
where
\begin{gather*}
V_0     =\sqrt{2}G_FN_0\simeq7.63\times10^{-14}\,\text{eV},
\quad
\left(L_0=\frac{2\pi}{V_0}\simeq1.62\times10^4\,\text{km}\right), \\
a_\tau   =\frac{3\alpha r_\tau\left[\ln(1/r_\tau)-1\right]}
          {4\pi\sin^2\theta_W}
          \simeq 2.44\times10^{-5},
\quad
b_\tau   =\frac{\ln(1/r_\tau)-2/3}{\ln(1/r_\tau)-1}
          \simeq 1.05,
\end{gather*}
$G_F$ is the Fermi coupling constant, $\alpha$ is the
fine-structure constant, $\theta_W$ is the weak-mixing angle
and $r_\tau=(m_\tau/m_W)^2$.
Thus, for an isoscalar medium $\left|q_R\right|$ is of the same
order of magnitude for any pair of flavors but $\nu_\mu-\nu_\tau$;
the ratio $q_R^{\left(\nu_\mu-\nu_\tau\right)}/
q_R^{\left(\nu_e-\nu_\mu\right)}$ is about $-5\times10^{-5}$.
For certain regions of a neutron-rich medium the value of
$q_R^{\left(\nu_e-\nu_s\right)}$ may become vanishingly small.
In this case, the one-loop radiative corrections must be taken
into account. For very high energies, all the expressions
\eqref{qR_Examples} have to be corrected for the gauge boson
propagators and strong-interaction effects.

One can expect $\left|q_R\right|$ to be either an
energy-independent or decreasing function for any pair of mixed
neutrino flavors. On the other hand, as noted above,
there are several cases of much current interest
when $\left|q_I\right|$ either increases with energy without bound
(mixing between active and sterile neutrino states) or has a broad
or sharp maximum (as for $\nu_\mu-\nu_\tau$ or
$\overline{\nu}_e-\overline{\nu}_\mu$ mixings, respectively).
Numerical estimations, performed using the results of refs.
\cite{NeutrinoInteractions} and  \cite{TauAppearance}, suggest that
for every of these cases there is an energy range in which $q_R$
and $q_I$ are comparable in magnitude.
Considering that both $q_R$ and $q_I$ are proportional to
the density and besides are dependent upon the composition
of the medium there may exist even more specific situations,
when $\left|q_R\right|\sim\left|q_I\right|\sim\left|\Delta\right|$
or $\left|q_R\right|\sim\left|\Delta_c\right|$ and
$\left|q_I\right|\sim\left|\Delta_s\right|$.
If this is the case, the refraction, absorption and mixing
become interestingly superimposed.

\protect\section{Eigenproblem and mixing matrix in matter}
\label{sec:Eigenproblem}

The matrix \eqref{NewH} has two complex instantaneous eigenvalues,
$\varepsilon(t)$ and $-\varepsilon(t)$, with
$\varepsilon=\varepsilon_R+i\varepsilon_I$
satisfying the characteristic equation
\begin{equation}\label{CharacteristicEquation}
\varepsilon^2=\left(q-q_+\right)\left(q-q_-\right),
\end{equation}
where $q_\pm=\Delta_c \pm i\Delta_s=\Delta e^{\pm2i\theta}$.
The solution to eq. \eqref{CharacteristicEquation} for
$\varepsilon_R$ and $\varepsilon_I$ may be written as
\begin{subequations}\label{Eigenvalue}
\begin{align}
\label{Eigenvalue_R^2}
\varepsilon_R^2&=\frac{1}{2}\left(\varepsilon_0^2-q_I^2\right)
          +\frac{1}{2}\sqrt{\left(\varepsilon_0^2-q_I^2\right)^2
                 +4q_I^2\left(\varepsilon_0^2-\Delta_s^2\right)}, \\
\label{Eigenvalue_I}
\varepsilon_I  &=\frac{q_I\left(q_R-\Delta_c\right)}
{\varepsilon_R} \quad\left(\text{provided $q_R\neq\Delta_c$}\right),
\end{align}
\end{subequations}
with
\[
\varepsilon_0=\sqrt{\Delta^2-2\Delta_cq_R+q_R^2} \geq
              \left|\Delta_s\right|.
\]
The sign of $\varepsilon_R$ is a matter of convention.
We define $\text{sign}\left(\varepsilon_R\right)=
\text{sign}(\Delta)\equiv\zeta$. At that choice
$\varepsilon=\Delta$ for vacuum ($q=0$) and
$\varepsilon=\zeta\varepsilon_0$ if $q_I=0$.

\begin{figure}[htb]
\center\includegraphics[width=0.6\linewidth]{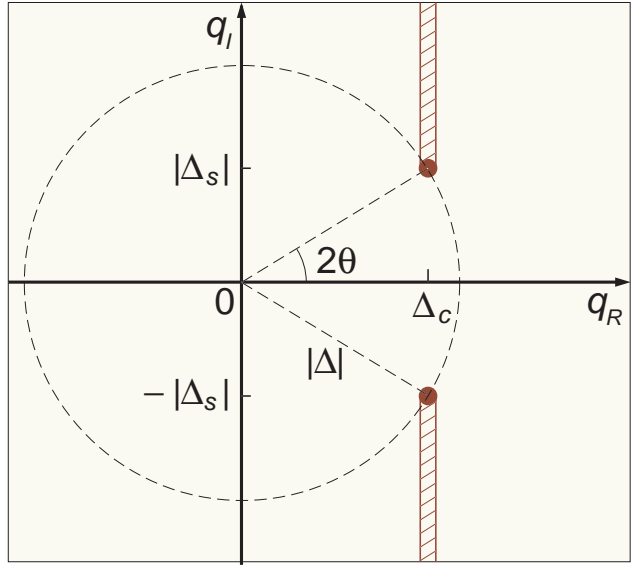}
\protect\caption{Zeros and cuts of $\varepsilon$ in the $q$
                 plane for $\Delta_c>0$.
\label{f:plane}}
\end{figure}

Let us consider the behavior of $\varepsilon_R$
and $\varepsilon_I$ in the vicinity of the MSW resonance,
defined by the condition $q_R=q_R(t_\star)=\Delta_c$
(we suppose for simplicity that the matter density and
composition are continuous functions of $t$).
The accurate passage to the limit in eqs. \eqref{Eigenvalue}
gives
\begin{align*}
\lim_{q_R\to\Delta_c\pm0}\varepsilon_R&=
\Delta_s\sqrt{\max\left(1-\Delta_I^2/\Delta_s^2,0\right)}, \\
\lim_{q_R\to\Delta_c\pm0}\varepsilon_I&=
\pm\zeta\Delta_I\sqrt{\max\left(1-\Delta_s^2/\Delta_I^2,0\right)},
\end{align*}
where $\Delta_I$ is taken to be the value of $q_I$ in the MSW point
(that is $\Delta_I=q_I(t_\star)$).
Therefore the resonance value of $\left|\varepsilon_R\right|$
(which is inversely proportional to the neutrino oscillation length
in matter) is always {\em smaller} than the conventional MSW value
$\left|\Delta_s\right|$ and {\em vanishes} if $\Delta_I^2<\Delta_s^2$
($\varepsilon_I$ remains finite in this case).
In neutrino transition through the region of resonance density
$\rho=\rho(t_\star)$, $\varepsilon_I$ undergoes discontinuous jump
while $\varepsilon_R$ remains continuous. The corresponding cuts in
the $q$ plane are placed outside the circle $|q|\leq|\Delta|$
as is shown in fig. \ref{f:plane}.
If $\Delta_I^2>\Delta_s^2$, the imaginary part of $\varepsilon$
vanishes while the real part is kept finite.

A distinctive feature of eq. \eqref{CharacteristicEquation}
is the existence of two mutually conjugate ``super-resonance''
points $q_\pm$ in which $\varepsilon$ vanishes giving rise to the
{\em total degeneracy} of the levels of the system under
consideration. Certainly, the behavior of the system in the vicinity
of these points must be dramatically different from the conventional
pattern. As noted in sec. \ref{sec:ME}, the ``super-resonance''
conditions are physically realizable for various meaningful
mixing scenarios.

In order to simplify the solution to the eigenstate problem we will
assume that the phase trajectory $q=q(t)$ does not cross the points
$q_\pm$ at any $t$. Within the framework of the non-Hermitian quantum
dynamics (see, e.g., ref. \cite{NHdynamics} and references therein)
one has to consider the two pairs of instantaneous eigenvectors
$\vert\varPsi_\pm\rangle$ and $\vert\overline{\varPsi}_\pm\rangle$
which obey the relations
\begin{equation}\label{EigenvectorProblem}
   \mathbf{H}     \vert\varPsi_\pm           \rangle=
\pm\,\varepsilon  \vert\varPsi_\pm           \rangle
\quad\text{and}\quad
    \mathbf{H}^\dag\vert\overline{\varPsi}_\pm\rangle=
\pm\,\varepsilon^*\vert\overline{\varPsi}_\pm\rangle.
\end{equation}
For $q\neq q_\pm$ these pairs form a complete biorthogonal
and biorthonormal set,
\[
\langle\overline{\varPsi}_\pm\vert\varPsi_\pm\rangle=1, \quad
\langle\overline{\varPsi}_\pm\vert\varPsi_\mp\rangle=0, \quad
 \vert\varPsi_+\rangle\langle\overline{\varPsi}_+\vert
+\vert\varPsi_-\rangle\langle\overline{\varPsi}_-\vert
=\mathbf{1}.
\]
Therefore, the eigenvectors are defined up to a gauge transformation
\[
\vert         {\varPsi}_\pm\rangle \mapsto e^{ if_\pm  }
\vert         {\varPsi}_\pm  \rangle,                       \quad
\vert\overline{\varPsi}_\pm\rangle \mapsto e^{-if_\pm^*}
\vert\overline{\varPsi}_\pm\rangle,
\]
with arbitrary complex functions $f_\pm(t)$ such that
$\text{Im}\left(f_\pm\right)$ vanish as $q=0$.%
\footnote{For our aims, the class of the gauge functions may be
restricted without loss of generality by the condition
$\left.f_\pm\right|_{q=0}=0$.}
Thus it is sufficient to find any particular solution of eqs.
\eqref{EigenvectorProblem}. Taking into account that
$\mathbf{H}^\dag=\mathbf{H}^*$, we may set
$\vert\overline{\varPsi}_\pm\rangle=\vert\varPsi_\pm^*\rangle$
and hence the eigenvectors can be found from the identity
\[
\mathbf{H}=
 \varepsilon\vert\varPsi_+\rangle\langle\varPsi_+^*\vert
-\varepsilon\vert\varPsi_-\rangle\langle\varPsi_-^*\vert.
\]
Setting
$\vert\varPsi_\pm\rangle=\left(v_\pm,\pm v_\mp\right)^T$
we arrive at the equations
\begin{equation}\label{Comp1}
v_\pm^2=\frac{\varepsilon\pm(q-\Delta_c)}{2\varepsilon}, \quad
v_+v_-=\frac{\Delta_s}{2\varepsilon},
\end{equation}
a particular solution of which can be written as
\begin{equation}\label{Comp2}
v_+=\sqrt{\left\lvert\frac{\varepsilon+q-\Delta_c}
{2\varepsilon}\right\rvert}\,e^{i(\varphi-\psi)/2}, \quad
v_-=\zeta\sqrt{\left\lvert\frac{\varepsilon-q+\Delta_c}
      {2\varepsilon}\right\rvert}\,e^{i(-\varphi-\psi)/2},
\end{equation}
where
\begin{gather*}
\varphi=\arg(\varepsilon+q-\Delta_c)=-\arg(\varepsilon-q+\Delta_c)
       =\arctan\left(\frac{q_I}{\varepsilon_R}\right), \\
\psi   =\arg(\varepsilon)
       =\arctan\left(\frac{\varepsilon_I}{\varepsilon_R}\right),
\end{gather*}
and we have fixed the remaining gauge ambiguity by a comparison
with the vacuum case.

It may be sometimes useful to define the complex mixing angle
in matter $\varTheta$ by the relations $\sin\varTheta=v_+$
and $\cos\varTheta=v_-$ or, equivalently,
\[
\sin2\varTheta=\frac{\Delta_s  }{\varepsilon}, \quad
\cos2\varTheta=\frac{\Delta_c-q}{\varepsilon}.
\]
The real and imaginary parts of $\varTheta$ are found
to be
\[
\begin{aligned}
\text{Re}(\varTheta)&=\frac{1}{2}\arctan\left[
\frac{\left(q_I-\Delta_s\right)\varepsilon_R
     -\left(q_R-\Delta_c\right)\varepsilon_I}
     {\left(q_R-\Delta_c\right)\varepsilon_R
     +\left(q_I-\Delta_s\right)\varepsilon_I}\right], \\
\text{Im}(\varTheta)&=\frac{1}{4}\ln\left[
\frac{\varepsilon_R^2+\varepsilon_I^2}
{\left(q_R-\Delta_c\right)^2+\left(q_I-\Delta_s\right)^2}\right].
\end{aligned}
\]
Having regard to the above-mentioned prescription for
the sign of $\varepsilon_R$, one can verify that
$\varTheta=\theta$ if $q=0$ (vacuum case) and
$\varTheta=0$ if $\Delta_s=0$ (no mixing or $m_1^2=m_2^2$).
It is also clear that $\varTheta$ becomes the standard MSW
mixing angle with $\text{Im}(\varTheta)=0$ when $q_I=0$
($\varLambda_\alpha=\varLambda_\beta$).

In order to build up the solution to eq. \eqref{NewWE} for the
nondegenerate case one has to diagonalize the Hamiltonian
$\mathbf{H}$. Generally a non-Hermitian matrix cannot be
diagonalized by a single unitary transformation. But in our
simple case this can be done by a complex orthogonal matrix
(extended mixing matrix in matter)
\[
\mathbf{U}_f=\mathbf{U}\exp(i\mathbf{f}),
\]
where $\mathbf{f}=\text{diag}\left(f_-,f_+\right)$, $f_\pm$ are
the arbitrary complex phases mentioned above and
\[
\mathbf{U}=\left(\vert\varPsi_-\rangle,\vert\varPsi_+\rangle\right)=
\begin{pmatrix}
 v_- & v_+ \\
-v_+ & v_-
\end{pmatrix}=
\begin{pmatrix}
 \cos\varTheta & \sin\varTheta \\
-\sin\varTheta & \cos\varTheta
\end{pmatrix}.
\]
The matrix $\mathbf{U}$ by its construction has the following
properties:
\begin{equation}\label{Uprop}
\mathbf{U}^T\mathbf{H}\mathbf{U}=
\text{diag}\left(-\varepsilon,\varepsilon\right), \quad
\mathbf{U}^T\mathbf{U}=\mathbf{1},                \quad
\left.\mathbf{U}\right|_{q=0}=\mathbf{V}.
\end{equation}
From eq. \eqref{CharacteristicEquation} it follows that
$\partial\varepsilon/\partial q=(q-\Delta_c)/\varepsilon$ and
thus eqs. \eqref{Comp1} yield
\[
\frac{\partial v_\pm}{\partial q}=
\pm \frac{\Delta_s^2v_\mp}{2\varepsilon^2}.
\]
We therefore have
\begin{equation}\label{NonAdiabaticTerm}
i\mathbf{U}^T\dot{\mathbf{U}}=-\Omega
\begin{pmatrix}
0 & -i \\
i &  0
\end{pmatrix}=
-\Omega\boldsymbol{\sigma}\!\strut_2,
\end{equation}
where
\[
\Omega=\frac{\dot{q}\Delta_s}{2\varepsilon^2}=
\frac{i}{4}\frac{\d}{\d t}\ln\left(\frac{q-q_+}{q-q_-}\right).
\]
According to eqs. \eqref{Uprop} and \eqref{NonAdiabaticTerm},
\begin{subequations}\label{Ufprop}
\begin{gather}
\label{Ufprop1}
\mathbf{U}_f^T\mathbf{H}\mathbf{U}_f=
\text{diag}\left(-\varepsilon,\varepsilon\right),  \quad
\mathbf{U}_f^T\mathbf{U}_f=\mathbf{1}, \quad
\left.\mathbf{U}_f\right|_{q=0}=\mathbf{V}, \\
\label{Ufprop2}
i\mathbf{U}_f^T\dot{\mathbf{U}}_f=
-\Omega e^{-i\mathbf{f}}\boldsymbol{\sigma}\!\strut_2
e^{i\mathbf{f}}-\dot{\mathbf{f}}.
\end{gather}
\end{subequations}

\protect\section{Adiabatic solution}
\label{sec:GS}

With the mixing matrix $\mathbf{U}_f$ in hand we can write
the formal solution to eq. \eqref{NewWE} in the most general
form:
\begin{equation}\label{MostGenForm}
\widetilde{\mathbf{S}}(t)=\mathbf{U}_f(t)
\exp\left[-i\boldsymbol{\Phi}(t)\right]
\mathbf{X}_f(t)\mathbf{U}_f^T(0).
\end{equation}
Here $\mathbf{X}_f(t)$ is an unknown matrix, $\boldsymbol{\Phi}(t)=
\text{diag}\left(-\varPhi(t),\varPhi(t)\right)$ and
$\varPhi(t)=\varPhi_R(t)+i\varPhi_I(t)$ is the complex
dynamical phase, defined by
\[
\varPhi_R(t)=\int_0^t\varepsilon_R(t')\d t', \quad
\varPhi_I(t)=\int_0^t\varepsilon_I(t')\d t'.
\]
Substituting eq. \eqref{MostGenForm} into eq. \eqref{NewWE} and
using eqs. \eqref{Ufprop}, we find
\[
i\dot{\mathbf{X}}_f(t)=\left[\Omega(t)e^{-i\mathbf{f}(t)}
\mathbf{F}(t)e^{i\mathbf{f}(t)}
+\dot{\mathbf{f}}(t)\right]\mathbf{X}_f(t),
\quad \mathbf{X}_f(0)=\mathbf{1},
\]
where
\[
\mathbf{F}(t)=e^{i\boldsymbol{\Phi}(t)}
\boldsymbol{\sigma}_2
e^{-i\boldsymbol{\Phi}(t)}=
\begin{pmatrix}
0 & -ie^{-2i\varPhi(t)} \\
ie^{2i\varPhi(t)} & 0
\end{pmatrix}.
\]
It can be proved now that the right side of eq. \eqref{MostGenForm}
is gauge-invariant i.e. it {\em does not depend} on the unphysical
complex phases $f_\pm(t)$. This crucial fact
is closely related to the absence of the Abelian topological
phases in the system under consideration.%
\footnote{Note that for $3\nu$ case the construction corresponding
          to eq. \eqref{MostGenForm} is not generally gauge-invariant
          (and thus is unphysical) even in the Hermitian case
          \cite{Naumov} due to the nontrivial (irremovable)
          topological phases. However, it becomes invariant both in
          the Hermitian \cite{Naumov} and non-Hermitian
          \cite{Korenblit95} cases after the substitution
          $\boldsymbol{\Phi}\mapsto\boldsymbol{\Phi}
          +\boldsymbol{\Gamma}$, with $\boldsymbol{\Gamma}$ a
          diagonal matrix incorporating the topological phases.}
Finally, we can put $f_\pm=0$ in eq. \eqref{MostGenForm} and the
result is
\begin{subequations}\label{Gen_Form}
\begin{gather}
\label{Gen_S}
\widetilde{\mathbf{S}}(t)=\mathbf{U}(t)
\exp\left[-i\boldsymbol{\Phi}(t)\right]
\mathbf{X}(t)\mathbf{U}^T(0), \\
\label{Eq_for_X}
i\dot{\mathbf{X}}(t)=\Omega(t)\mathbf{F}(t)\mathbf{X}(t),
\quad \mathbf{X}(0)=\mathbf{1}.
\end{gather}
\end{subequations}

These equations, being equivalent to eq. \eqref{NewWE},
have nevertheless a restricted range of practical usage
on account of poles and cuts as well as decaying and increasing
exponents in the ``Hamiltonian'' $\Omega\mathbf{F}$. However, the
adiabatic theorem of Hermitian quantum mechanics (see, e.g., ref.
\cite{Messiah75}) can straightforwardly be extended to
eq. \eqref{NewWE} under the following requirements:
\begin{description}
\item[\rm{(a)}] the potential $q$ is a sufficiently smooth and slow
                function of $t$;
\item[\rm{(b)}] the imaginary part of the dynamical phase is a bounded
                function i.e.
                $\lim_{t\to\infty}\left|\varPhi_I(t)\right|$
                is finite;
\item[\rm{(c)}] the phase trajectory $q=q(t)$ is placed far from the
                singularities for any $t$.
\end{description}

The first requirement breaks down for a condensed medium with a sharp
boundary or layered structure (like the Earth). If however the
requirement (a) is valid inside each layer
$\left(t_i,t_{i+1}\right)$, the problem reduces to
eqs. \eqref{Gen_Form} by applying the rule
\[
\widetilde{\mathbf{S}}(t)\equiv
\widetilde{\mathbf{S}}(t,0)=
\widetilde{\mathbf{S}}\left(t,t_n\right)\ldots
\widetilde{\mathbf{S}}\left(t_2,t_1\right)
\widetilde{\mathbf{S}}\left(t_1,0\right),
\]
where $\widetilde{\mathbf{S}}\left(t_{i+1},t_i\right)$ is the
time-evolution operator for the $i$-th layer.

The requirement (b) alone is not too restrictive considering that
for many astrophysical objects (like stars, galactic nuclei, jets
and so on) the density $\rho$ exponentially disappears to the
periphery and, on the other hand, $\varepsilon_I\to0$ as
$\rho\to0$. In this instance, the function $\varPhi_I(t)$ must be $t$
independent for sufficiently large $t$. But, in the case of a steep
density profile, the requirements (a) and (b) may be inconsistent.

The important case of violation of the requirement (c) is the
subject of a special study which is beyond the scope of this paper.
It is interesting to note in this connection that, in the Hermitian
case, a general adiabatic theorem has been proved without the
traditional gap condition \cite{Avron99}.

Having regard to the mentioned limitations, we can expect that
eqs. \eqref{Gen_Form} is tolerable for finding the approximate,
adiabatic solution to the problem (and the corrections to
that solution) for many cases of pragmatic interest.
We shall presume now that the parameters of the matter vary so
slowly that all necessary conditions do hold for $0\leq t \leq T$.
Then, in the adiabatic limit, we can put $\Omega=0$ in
eq. \eqref{Eq_for_X}. Therefore $\mathbf{X}=\mathbf{1}$ and
eq. \eqref{Gen_S} yields
\begin{align*}
\widetilde{S}_{\alpha\alpha}(t)&=
         v_+(0)v_+(t)e^{-i\varPhi(t)}+v_-(0)v_-(t)e^{i\varPhi(t)}, \\
\widetilde{S}_{\alpha\beta }(t)&=
         v_-(0)v_+(t)e^{-i\varPhi(t)}-v_+(0)v_-(t)e^{i\varPhi(t)}, \\
\widetilde{S}_{\beta \alpha}(t)&=
         v_+(0)v_-(t)e^{-i\varPhi(t)}-v_-(0)v_+(t)e^{i\varPhi(t)}, \\
\widetilde{S}_{\beta \beta }(t)&=
         v_-(0)v_-(t)e^{-i\varPhi(t)}+v_+(0)v_+(t)e^{i\varPhi(t)}.
\end{align*}
Taking into account eq. \eqref{ProbNew} we obtain the
survival and transition probabilities:
\begin{equation}\label{AdiabaticSolution}
\begin{aligned}
P_{\alpha\alpha}(t)&=A(t)\left\{
 \left[I_+^+(t)e^{\varPhi_I(t)}+I_-^-(t)e^{-\varPhi_I(t)}\right]^2
     -I^2(t)\sin^2\left[\varPhi_R(t)-\varphi_+(t)\right]\right\}, \\
P_{\alpha\beta }(t)&=A(t)\left\{
 \left[I_+^-(t)e^{\varPhi_I(t)}-I_-^+(t)e^{-\varPhi_I(t)}\right]^2
     +I^2(t)\sin^2\left[\varPhi_R(t)-\varphi_-(t)\right]\right\}, \\
P_{\beta \alpha}(t)&=A(t)\left\{
 \left[I_-^+(t)e^{\varPhi_I(t)}-I_+^-(t)e^{-\varPhi_I(t)}\right]^2
     +I^2(t)\sin^2\left[\varPhi_R(t)+\varphi_-(t)\right]\right\}, \\
P_{\beta \beta }(t)&=A(t)\left\{
 \left[I_-^-(t)e^{\varPhi_I(t)}+I_+^+(t)e^{-\varPhi_I(t)}\right]^2
     -I^2(t)\sin^2\left[\varPhi_R(t)+\varphi_+(t)\right]\right\},
\end{aligned}
\end{equation}
where we have denoted for compactness
\begin{gather*}
I_{\varsigma}^{\varsigma'}(t)=
\left|v_{\varsigma}(0)v_{\varsigma'}(t)\right|,\quad
\varsigma,\varsigma'=\pm,                                        \\
I^2(t)=4I_+^+(t)I_-^-(t)=4I_+^-(t)I_-^+(t)=
\frac{\Delta_s^2}{\left|\varepsilon(0)\varepsilon(t)\right|},    \\
\varphi_\pm(t)=\frac{1}{2}\left[\varphi(0)\pm\varphi(t)\right].
\end{gather*}

In the event that the conditions
\begin{subequations}\label{MSWconditions}
\begin{gather}
\label{MSWcondition_1}
\left|\frac{1}{\varLambda_\beta (t)}
     -\frac{1}{\varLambda_\alpha(t)}\right|\ll4\varepsilon_0(t) \\
\intertext{and}
\label{MSWcondition_2}
t\ll\min\left[\varLambda_\alpha(t),\varLambda_\beta(t)\right]
\end{gather}
\end{subequations}
are satisfied for any $t\in[0,T]$, the formulas
\eqref{AdiabaticSolution} reduce to the standard MSW
adiabatic solution
\begin{equation}\label{MSWLimit}
\left.
\begin{aligned}
P_{\alpha\alpha}(t)&=P_{\beta\beta }(t)=
                     \frac{1}{2}\left[1+J(t)\right]
                     -I_0^2(t)\sin^2\left[\varPhi_0(t)\right], \\
P_{\alpha\beta }(t)&=P_{\beta\alpha}(t)=
                     \frac{1}{2}\left[1-J(t)\right]
                     +I_0^2(t)\sin^2\left[\varPhi_0(t)\right], \\
\end{aligned}
\right\}\qquad\text{(MSW)}
\end{equation}
where
\begin{gather*}
J(t)=\frac{\Delta^2-\Delta_c\left[q_R(0)+q_R(t)\right]+q_R(0)q_R(t)}
{\varepsilon_0(0)\varepsilon_0(t)}, \\
I_0^2(t)=\frac{\Delta_s^2}{\varepsilon_0(0)\varepsilon_0(t)}, \quad
\varPhi_0(t)=\int_0^t\varepsilon_0(t')\d t'.
\end{gather*}
Needless to say either of the conditions \eqref{MSWconditions} or
both may be violated for sufficiently high neutrino energies and/or
for thick media, resulting in radical differences between the two
solutions. These differences are of obvious interest to
high-energy neutrino astrophysics.

It is perhaps even more instructive to examine the distinctions
between the general adiabatic solution
\eqref{AdiabaticSolution} and its ``classical limit''%
\footnote{Considering that $\Omega\propto\Delta_s$, the solution
         \eqref{ClassicalLimit} is exact (for $\Delta_s=0$) and can
         be derived directly from eq. \eqref{NewWE}. To make
         certain that the adiabatic solution has correct
         classical limit, the following relations are useful:
          \[
          \lim_{\Delta_s\to0}\varepsilon(t)=
          \zeta\zeta_R\left[q(t)-\Delta_c\right]
          \quad\text{and}\quad
          \lim_{\Delta_s\to0}\left|v_\pm\right(t)|^2=
          \frac{1}{2}\left(\zeta\zeta_R\pm1\right),
          \]
          where $\zeta_R=\text{sign}\left[q_R(t)-\Delta_c\right]$.}
\begin{equation}\label{ClassicalLimit}
\left.
\begin{aligned}
P_{\alpha\alpha}(t)&=
\exp\left[-\int_0^t\frac{\d t'}{\varLambda_\alpha(t')}\right], \quad
P_{\alpha\beta }(t) =0,                                        \\
P_{\beta \beta }(t)&=
\exp\left[-\int_0^t\frac{\d t'}{\varLambda_\beta (t')}\right], \quad
P_{\beta \alpha}(t)=0,
\end{aligned}
\right\}\qquad\text{($\Delta_s=0$)}
\end{equation}
which takes place either in the absence of mixing or for
$m_1^2=m_2^2$.

In the next section we consider some features of the solution
\eqref{AdiabaticSolution} for media with constant density and
composition ($\dot{q}\equiv0$).
For this simple case, the adiabatic approximation becomes exact
and thus free from the above-mentioned conceptual difficulties.

\protect\section{Matter of constant density and composition}
\label{sec:MCD}

For definiteness sake we assume $\varLambda_\alpha<\varLambda_\beta$
(and thus $q_I<0$) from here. The opposite case can be considered in
a similar way. Let us denote
\begin{gather*}
\frac{1}{\varLambda_\pm}=
\frac{  1   }{2}\left(\frac{1}{\varLambda_\alpha}
                     +\frac{1}{\varLambda_\beta}\right)
\pm\frac{\xi}{2}\left(\frac{1}{\varLambda_\alpha}
                     -\frac{1}{\varLambda_\beta}\right), \\
I_\pm^2=\frac{1}{4}\left(1+\frac{\varepsilon_0^2+q_I^2-\Delta_s^2}
        {\varepsilon_R^2+\varepsilon_I^2}\right) \pm
  \frac{\xi}{2}\left(\frac{\varepsilon_R^2+q_I^2}
        {\varepsilon_R^2+\varepsilon_I^2}\right), \\
L  =\frac{\pi}{\left|\varepsilon_R\right|}
\quad\text{and}\quad
\xi=\left|\frac{q_R-\Delta_c}{\varepsilon_R}\right|.
\end{gather*}
As is easy to see,
\[
I_\pm^\pm=
\begin{cases}
I_\pm & \text{if}\quad\text{sign}\left(q_R-\Delta_c\right)=+\zeta, \\
I_\mp & \text{if}\quad\text{sign}\left(q_R-\Delta_c\right)=-\zeta,
\end{cases}
\]
\[
I_+^-=I_-^+=\sqrt{I_+I_-}=\frac{I}{2}=
\left|\frac{\Delta_s}{2\varepsilon}\right|
\quad\text{and}\quad
\text{sign}(\varphi)=-\zeta.
\]
By applying these identities and formulas from sec. \ref{sec:GS},
the neutrino oscillation probabilities can be written as
\begin{subequations}\label{ConstantDensitySolution}
\begin{gather}
P_{\alpha\alpha}(t)=
\left(I_+e^{-t/2\varLambda_+}+I_-e^{-t/2\varLambda_-}\right)^2
-I^2e^{-t/\varLambda}\sin^2\left(\frac{\pi t}{L}+|\varphi|\right),\\
P_{\beta \beta }(t)=
\left(I_-e^{-t/2\varLambda_+}+I_+e^{-t/2\varLambda_-}\right)^2
-I^2e^{-t/\varLambda}\sin^2\left(\frac{\pi t}{L}-|\varphi|\right),\\
P_{\alpha\beta }(t)=P_{\beta \alpha}(t)=\frac{1}{4}I^2
 \left(e^{-t/2\varLambda_-}-e^{-t/2\varLambda_+}\right)^2
+I^2e^{-t/\varLambda}\sin^2\left(\frac{\pi t}{L}\right).
\end{gather}
\end{subequations}

\protect\subsection{Case $|q|\gtrsim\left|\Delta_s\right|$}
\label{ss:Case1}

Let us examine the case when $\varLambda_+$ and $\varLambda_-$
are vastly different in magnitude. This will be true when
$\varLambda_\beta\gg\varLambda_\alpha$ and the factor $\xi$ is
not too small. The second condition holds if $q_R$ is away from
the MSW resonance value $\Delta_c$ and the following dimensionless
parameter
\[
\varkappa=\frac{\Delta_s}{|q|}\approx0.033\times\sin2\theta
          \left(\frac{\Delta m^2}
          {10^{-3}\,\text{eV}^2}\right)
          \left(\frac{100\,\text{GeV}}{E_\nu}\right)
          \left(\frac{V_0}{|q|}\right)
\]
is sufficiently small. In fact we assume $|\varkappa|\lesssim1$
and impose no specific restriction for the ratio $q_R/q_I$.
The assumption spans several possibilities:
  (i) small $\Delta m^2$,
 (ii) small mixing angle,
(iii) high energy,
 (iv) high matter density.
The last two possibilities are of special interest because the
inequality $|\varkappa|\lesssim1$ may be fulfilled for a wide range of
the mixing parameters $\Delta m^2$ and $\theta$ by changing $E_\nu$
and/or $\rho$. In other words, this condition is by no means
artificial or too restrictive.

After simple while a bit tedious calculations we obtain
\begin{gather*}
\xi=1-\frac{1}{2}\varkappa^2+\mathcal{O}\left(\varkappa^3\right),
                                                                \quad
I^2=\varkappa^2+\mathcal{O}\left(\varkappa^3\right),               \\
I_+=1+\mathcal{O}\left(\varkappa^2\right),                      \quad
I_-=\frac{1}{4}\varkappa^2+\mathcal{O}\left(\varkappa^3\right).
\end{gather*}
Since $\varLambda_\alpha$ has been assumed to be small compared
to $\varLambda_\beta$, we have
\[
\varLambda\approx2\varLambda_\alpha,
\quad
\varLambda_+\approx\left(1+\frac{\varkappa^2}{4}\right)
            \varLambda_\alpha
            \approx\varLambda_\alpha, \quad
\varLambda_-\approx\left(\frac{4}{\varkappa^2}\right)
            \varLambda_\alpha
            \gg \varLambda_\alpha.
\]
In general the oscillation length in matter $L$ has no
Taylor expansion over the parameter $\varkappa$ and may
vary within a broad range depending on other circumstances.
For example, $L\approx\pi/|q|$ if $|q|\gg|\Delta|$
and $L\approx\pi/\sqrt{\varepsilon_0^2-\Delta_s^2}$ if
$|q_I|\gg\varepsilon_0>|\Delta_s|$.
Due to the wide spread among the length/time scales $\varLambda_\pm$,
$\varLambda$ and $L$ as well as among the amplitudes $I_\pm$ and $I$,
the regimes of neutrino oscillations are quite diverse for different
ranges of variable $t$.

\begin{figure}[t]
\center\includegraphics[width=\linewidth]{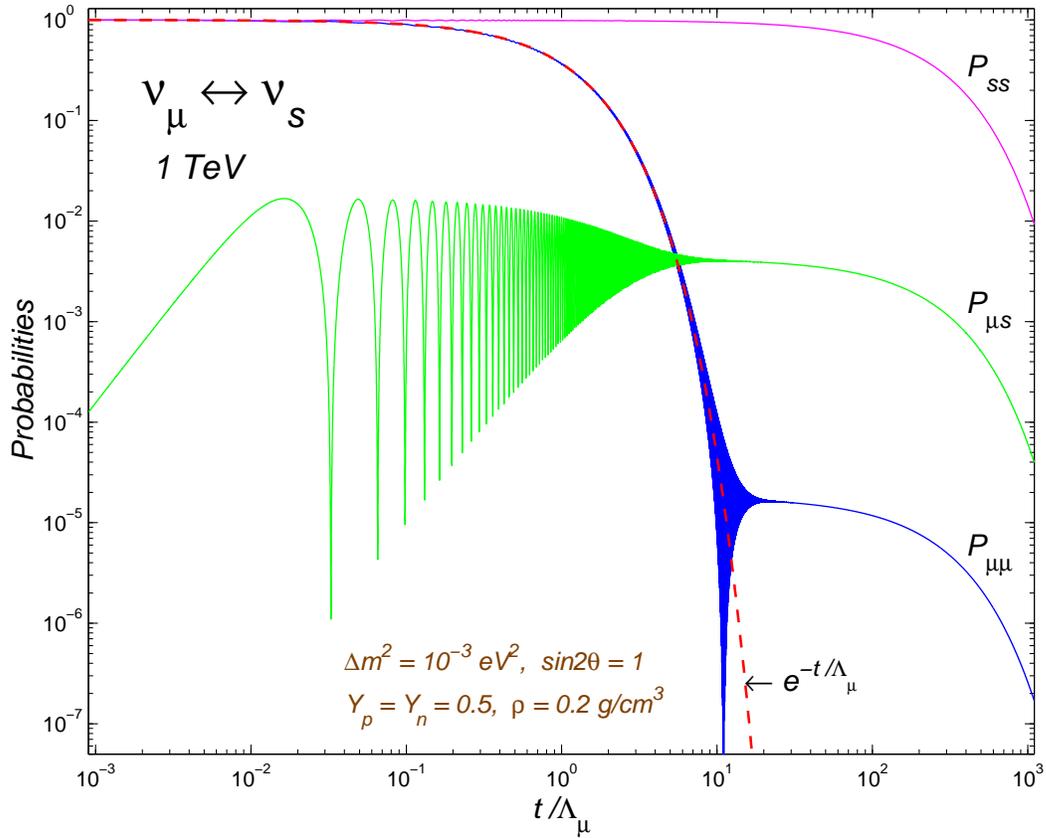}
\protect\caption{Survival and transition probabilities for the
                 $\nu_\mu\leftrightarrow\nu_s$ oscillations at
                 $E_\nu=1$ TeV and $\rho=0.2$ g/cm${}^3$.
                 The penetration coefficient
                 $\exp\left(-t/\varLambda_\mu\right)$ for
                 the case of unmixed muon neutrinos is shown by
                 dashed curve.
\label{f:1TeV}}
\end{figure}

\begin{figure}[t]
\center       \includegraphics[width=0.90\linewidth]{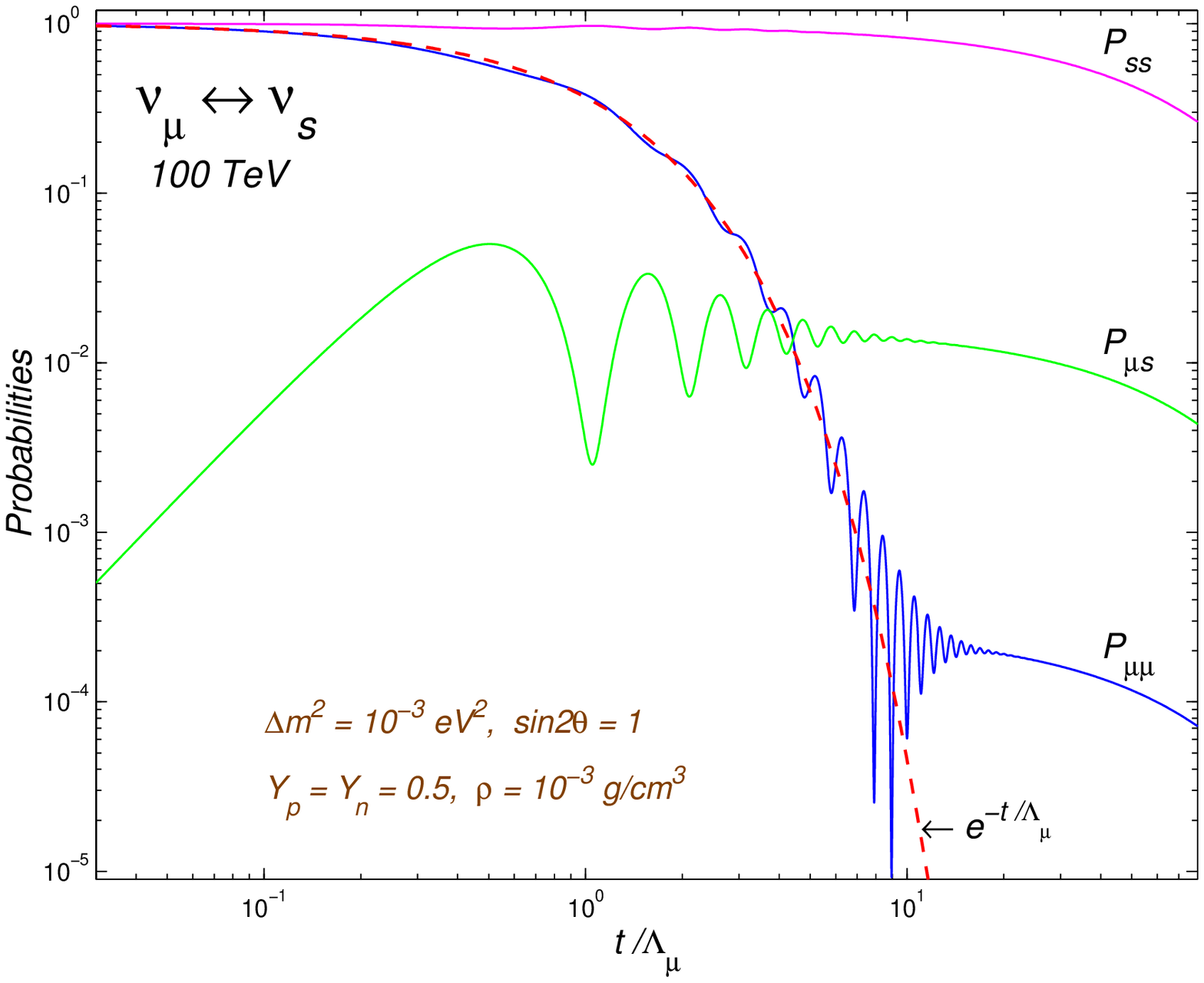}
\vspace*{-2mm}\includegraphics[width=0.90\linewidth]{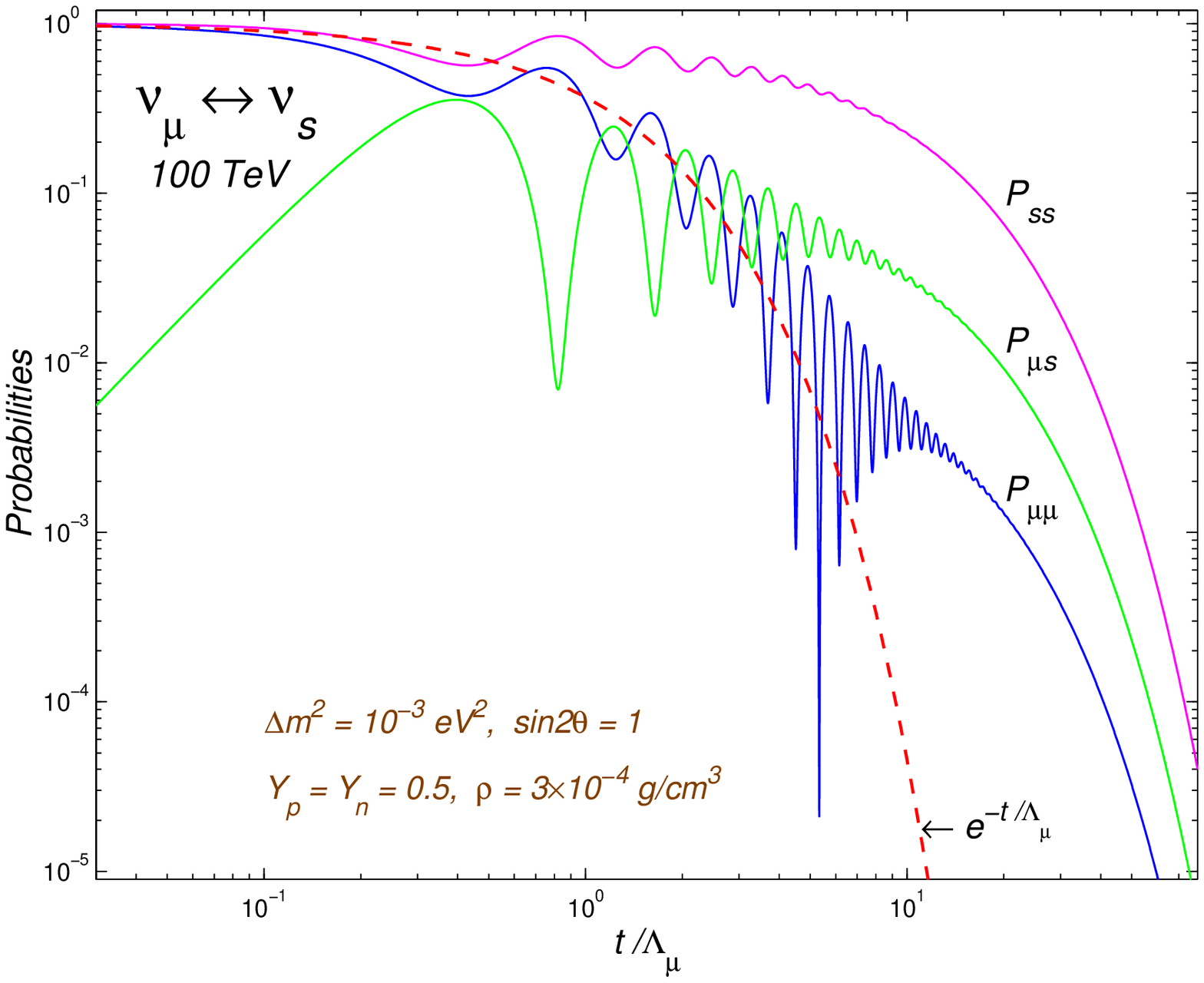}
\protect\caption{The same as in fig. \protect\ref{f:1TeV} but for
                 $E_\nu=100$ TeV and for two values of the matter
                 density $\rho$:
                 $10^{-3}$ g/cm${}^3$ (upper panel) and
                 $3\times10^{-4}$ g/cm${}^3$ (lower panel).
\label{f:100TeV}}
\end{figure}

Some of the essential features are illustrated in figs. \ref{f:1TeV}
and \ref{f:100TeV} by the example of $\nu_\mu-\nu_s$ oscillations in
an isoscalar medium ($Y_p=Y_n=0.5$). For such a medium the picture of
the $\nu_e-\nu_s$ ($\nu_\tau-\nu_s$) oscillations is the same
(nearly the same).
In all examples, the mixing parameters are taken to be
$\Delta m^2=10^{-3}$ eV${}^2$ and $\theta=\pi/4$.
In order to estimate the charged and neutral current $\nu_\mu N$
total cross sections we apply the results of ref.
\cite{NeutrinoInteractions} based on the CTEQ4-DIS parton
distributions \cite{Lai97}. The $\nu_\mu e$ contribution is
neglected.
Figure \ref{f:1TeV} depicts the survival and
transition probabilities for $E_\nu=1$ TeV and
$\rho=0.2\,\text{g/cm}^3$
($\varkappa\approx0.131$ in this case).
The classical limit of $P_{\mu\mu}$ (that is the penetration
coefficient for the unmixed muon neutrino flux) is also shown for
comparison.
In fig. \ref{f:100TeV} we present the same probabilities but for
$E_\nu=100$ TeV and (in order to demonstrate the density impact)
for the two values of $\rho$,
$10^{-3}\,\text{g/cm}^3$ (upper panel) and
$3\times10^{-4}\,\text{g/cm}^3$ (lower panel).
In these cases $\varkappa\approx0.261$ and 0.870, respectively.
With reference to figs. \ref{f:1TeV} and \ref{f:100TeV}, one can
see a regular gradation from slow (at $t\lesssim\varLambda_\mu$)
to very fast (at $t\gtrsim\varLambda_\mu$) neutrino oscillations
followed by the asymptotic nonoscillatory behavior
\clearpage
\clearpage
\[
P_{\mu\mu}(t)\simeq\frac{\varkappa^4}{16}e^{-t/\varLambda_-},
\quad
P_{s  s  }(t)\simeq e^{-t/\varLambda_-},
\quad
P_{\mu s}(t)=P_{s \mu}(t)\simeq
                   \frac{\varkappa^2}{4}e^{-t/\varLambda_-}.
\]
The latter regime is the most remarkable.
Since $\varLambda_-\gg\varLambda_\mu$, a small part of the flux
of muon neutrinos escapes through the medium of such a huge
thickness that would be absolutely opaque for them in the absence
of mixing.
At $t>(10-20)\varLambda_\mu$ the deviation of the survival
probability $P_{\mu\mu}$ from the classical expectation grows
exponentially.
The asymptotic value of the transition probability $P_{s\mu}$ is a
factor of $4/\varkappa^2$ in excess of $P_{\mu\mu}$.
This may be potentially beneficial for the observational neutrino
astrophysics, considering that after leaving the medium surrounding
an astrophysical neutrino source, the sterile neutrinos may travel
a very long distance in vacuum giving rise (through the vacuum
oscillations) to a detectable amount of muon neutrinos.

\protect\subsection{Degenerate case}
\label{ss:Case2}

\begin{figure}[t]
\center\includegraphics[width=\linewidth]{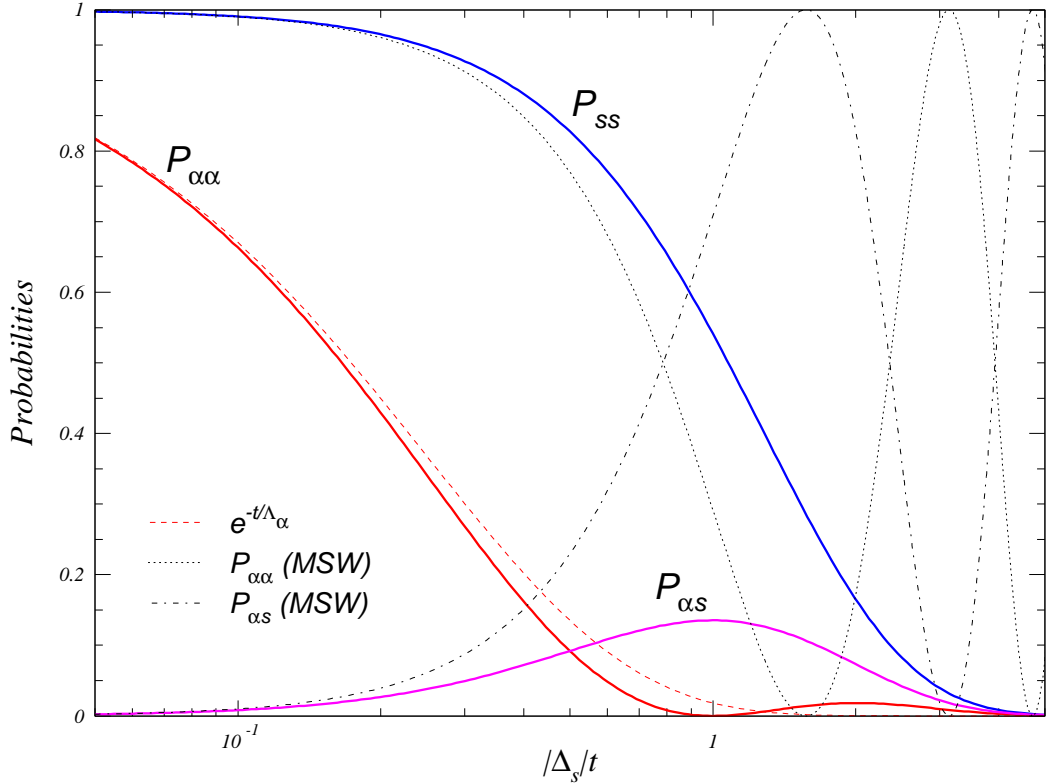}
\protect\caption{Survival and transition probabilities
                 for $\nu_\alpha\leftrightarrow\nu_s$ oscillations
                 in the case of degeneracy ($q=q_{-\zeta}$).
                 The standard MSW probabilities
                 (dotted and dash-dotted curves) together with the
                 penetration coefficient for unmixed $\nu_\alpha$
                 (dashed curve) are also shown.
\label{f:sr}}
\end{figure}

Our consideration must be completed for the case of degeneracy.
Due to the condition $q_I<0$, the density and composition of the
``degenerate environment'' are fine-tuned in such a way that
$q=q_{-\zeta}=\Delta_c-i\left|\Delta_s\right|$. The corresponding
formulas can be derived by a formal passage to the limit in
eqs. \eqref{ConstantDensitySolution}. A more simple way is
however in coming back to the master equation \eqref{NewWE}.
Indeed, in the limit of $q=q_{-\zeta}$, the Hamiltonian
\eqref{NewH} reduces to
\[
\mathbf{H}=\left|\Delta_s\right|
\begin{pmatrix}
 -i & \zeta \\
\zeta & i
\end{pmatrix}\equiv\left|\Delta_s\right|\mathbf{h}_\zeta.
\]
Considering that $\mathbf{h}_\zeta^2=\mathbf{0}$, we promptly
arrive at the solution of eq. \eqref{NewWE}:
\[
\widetilde{\mathbf{S}}(t)=
\mathbf{1}-it\left|\Delta_s\right|\mathbf{h}_\zeta
\]
and thus, taking into account eq. \eqref{ProbNew}, we have
\begin{gather*}
P_{\alpha\alpha}(t)=
      \left(1-\left|\Delta_s\right|t\right)^2e^{-t/\varLambda}, \\
P_{\beta \beta }(t)=
      \left(1+\left|\Delta_s\right|t\right)^2e^{-t/\varLambda}, \\
P_{\alpha\beta }(t)=P_{\beta \alpha}(t)=
                        \left(\Delta_st\right)^2e^{-t/\varLambda}.
\end{gather*}
Since
$1/\varLambda_\beta=1/\varLambda_\alpha-4\left|\Delta_s\right|$,
the necessary condition for the total degeneration is
$4\varLambda_\alpha\left|\Delta_s\right|\leq1$ and thus
$1/\varLambda=1/\varLambda_\alpha-2\left|\Delta_s\right|\geq
2\left|\Delta_s\right|$.
The equality only occurs when $\nu_\beta$ is sterile. Figure
\ref{f:sr} illustrates just this particular case. For comparison,
the standard MSW solution (see eq. \eqref{MSWLimit})
\[
\left.
\begin{aligned}
P_{\alpha\alpha}(t)=P_{ss}(t)=
\frac{1}{2}\left[1+\cos\left(2\Delta_st\right)\right], \\
P_{\alpha s}(t)=P_{s\alpha}(t)=
\frac{1}{2}\left[1-\cos\left(2\Delta_st\right)\right],
\end{aligned}
\right\}\qquad\text{(MSW)}
\]
is also shown as well as the classical penetration coefficient
$\exp\left(-t/\varLambda_\alpha\right)$ (with $1/\varLambda_\alpha$
{\em numerically} equal to $4\left|\Delta_s\right|$) relevant to
the transport of unmixed active neutrinos through the same
environment.

\protect\section{Conclusions}
\label{sec:Conclusions}

We have considered, on the basis of the MSW evolution equation with
complex indices of refraction, the conjoint effects of neutrino
mixing, refraction and absorption on high-energy neutrino propagation
through matter. The adiabatic solution with correct asymptotics in
the standard MSW and classical limits has been derived.
In the general case the adiabatic behavior is very different from
the conventional limiting cases.

A noteworthy example is given by the active-to-sterile neutrino
mixing. It has been demonstrated that, under proper conditions,
the survival probability of active neutrinos propagating through
a very thick medium of constant density may become many orders of
magnitude larger than it would be in the absence of mixing.
The quantitative characteristics of this phenomenon are highly
responsive to changes in density and composition of the
medium as well as to neutrino energy and mixing parameters.
Considering a great variety of latent astrophysical sources of
high-energy neutrinos, the effect may open a new window for
observational neutrino astrophysics.

\ack
It is a pleasure to thank Denis Comelli, Alexander Dolgov, Gianni
Fiorentini, Jambul Gegelia and Alexander Rudavets for useful
discussions.

\end{document}